\documentclass[aps,prx,reprint,groupedaddress,longbibliography]{revtex4-1}

\usepackage{amsmath,amsfonts}
\usepackage{graphicx}
\usepackage{hyperref}

\hypersetup{colorlinks,citecolor=blue,linkcolor=blue,urlcolor=blue}

\newcommand{\ket}[1]{\ensuremath{|#1\rangle}}

\newcommand{\rket}[1]{\ensuremath{\left|#1\right\rangle}}

\newcommand{\bra}[1]{\ensuremath{\left\langle#1\right|}}

\begin{document}

\title{Computational inverse method for constructing spaces of quantum models from wave functions}

\author{Eli Chertkov}
\affiliation{Department of Physics, University of Illinois at Urbana-Champaign, Urbana, Illinois 61801, USA}
\author{Bryan K. Clark}
\affiliation{Department of Physics, University of Illinois at Urbana-Champaign, Urbana, Illinois 61801, USA}

\begin{abstract}
Traditional computational methods for studying quantum many-body systems are ``forward methods,'' which take quantum models, i.e., Hamiltonians, as input and produce ground states as output. However, such forward methods often limit one's perspective to a small fraction of the space of possible Hamiltonians.
We introduce an alternative computational ``inverse method,'' the Eigenstate-to-Hamiltonian Construction (EHC), that allows us to better understand the vast space of quantum models describing strongly correlated systems. EHC takes as input a wave function $|\psi_T\rangle$ and produces as output Hamiltonians for which $|\psi_T\rangle$ is an eigenstate.  This is accomplished by computing the quantum covariance matrix, a quantum mechanical generalization of a classical covariance matrix. 
EHC is widely applicable to a number of models and in this work we consider seven different examples.
Using the EHC method, we construct a parent Hamiltonian with a new type of antiferromagnetic ground state, a parent Hamiltonian with two different targeted degenerate ground states, and large classes of parent Hamiltonians with the same ground states as well-known quantum models, such as the Majumdar-Ghosh model, the XX chain, the Heisenberg chain, the Kitaev chain, and a 2D BdG model.
EHC gives an alternative inverse approach for studying quantum many-body phenomena. 
\end{abstract}

\maketitle

\section{Introduction}

\begin{figure}
\begin{center}
\includegraphics[width=0.5\textwidth]{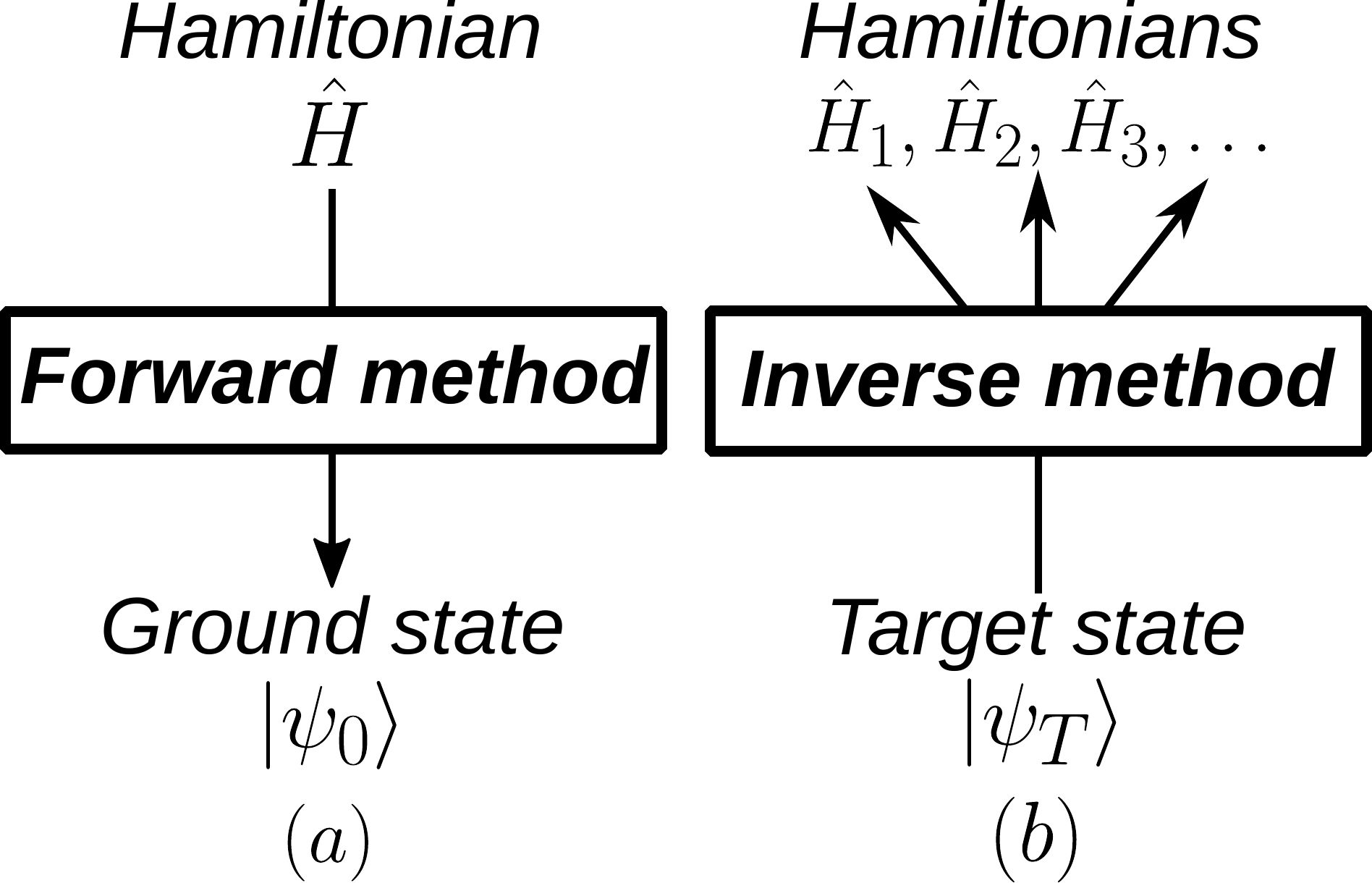}
\end{center}
\caption{(a) A typical \emph{forward method} used in quantum mechanics finds the ground state $\ket{\psi_0}$ of a single Hamiltonian $\hat{H}$. (b) We introduce a new \emph{inverse method}, EHC, that finds Hamiltonian(s) $\hat{H}_1,\hat{H}_2,\ldots$ from a target state $\ket{\psi_T}$, with the property that $\ket{\psi_T}$ is an energy eigenstate of these Hamiltonian(s).} \label{fig:ForwardInverseMethod}
\end{figure}

Our understanding of quantum many-body physics comes primarily from the use of ``forward methods.'' In the forward method approach, shown in Fig.~\ref{fig:ForwardInverseMethod}(a), a quantum model describing a material, e.g., a model Hamiltonian, is solved. Often solving each Hamiltonian is difficult requiring expensive numerics or complex analytic approaches. This restricts our attention to a few representative Hamiltonians or materials which support particular properties or interesting physics. However, the space of quantum models is vast and high-dimensional. The forward approach provides a limited perspective by restricting our focus to a small fraction of this space. The entire space, though, almost certainly contains a myriad of interesting physical Hamiltonians corresponding to undiscovered phases, unknown exactly solvable points, and Hamiltonians with desirable properties. 

While determining the ground state properties from a Hamiltonian is difficult, understanding interesting physics from simple prototypical wave functions is more straightforward. For this reason, wave functions such as resonating valence bond (RVB) states \cite{Anderson1973,Anderson1987}, projected BCS states \cite{Anderson1987,Gros1988,Paramekanti2001}, and Laughlin wave functions \cite{Laughlin1983,Trugman1985} have been widely used to understand spin liquids, high temperature superconductivity, and fractional quantum Hall physics in situations where Hamiltonian methods have not been feasible. Since these prototypical wave functions are easier to work with, one can consider using them as inputs for an ``inverse method'' approach for constructing parent Hamiltonians that have these wave functions as ground states. In fact, parent Hamiltonians have already been constructed in a variety of contexts and include, among others \cite{Wang2017,Kapit2010,PerezGarciaII2007,Gottesman1997,Raussendorf2003,Hein2004,Schroeter2007,Thomale2009,Greiter2014,Pekker2017}, RVB parent Hamiltonians on a Kagome lattice \cite{Seidel2009,Schuch2012,Zhou2014}, matrix product state parent Hamiltonians for one-dimensional systems \cite{Fannes1992,Nachtergaele1996,PerezGarciaI2007}, and Haldane pseudopotentials for a 2D electron gas \cite{Haldane1983,Stormer1999}. However, the methods for constructing parent Hamiltonians are wave function-specific, normally produce one or a small number of parent Hamiltonians, and often result in unphysical models. To overcome these limitations, we developed a novel inverse method that automates the construction of parent Hamiltonians from wave functions by searching for models in a large space of ``physically reasonable'' Hamiltonians. More broadly, inverse methods have been successful in applications such as solving machine learning problems \cite{LeCun2015}, targeting many-particle ordering in classical materials \cite{DiStasio2013,Marcotte2013,Chertkov2016,inverseopt,torquato_honeycomb,torquato_triangularsquarehoneycomb,cohnkumar,torquato_monotonicsquarehoneycomb,torquato_diamond,muller,hannon,zhang,jain1,jain2,negativethermalexpansion,negativepoissonratio}, and promoting certain properties in quantum many-body systems \cite{Ramezanpour2016,Changlani2015,Zheng2017,zunger1,zunger2,zunger3}.

Our new inverse method, Eigenstate-to-Hamiltonian Construction (EHC), takes as input a target wave function and a target space of Hamiltonians and produces as output the Hamiltonians within the target space for which the wave function is an eigenstate (see Fig.~\ref{fig:ForwardInverseMethod}(b)). EHC can be readily implemented with existing numerical tools. The key step of the method is the evaluation and analysis of the quantum covariance matrix (QCM) (see Eq.~(\ref{eq:CT})).  The wave functions provided as input to EHC need only to be represented in a way in which the QCM can be determined, such as numerically through the density matrix renormalization group (DMRG) \cite{White1992} or variational Monte Carlo (VMC) \cite{vmc1965}.  We show that EHC can be implemented in an efficient manner, with the procedure scaling quadratically in the number of variational parameters in the target space of Hamiltonians being considered. 

EHC helps solve an important general problem that has been actively pursued for decades, finding Hamiltonians with interesting ground state physics, by using the inverse approach of constructing parent Hamiltonians from wave functions. As described above, parent Hamiltonians have been constructed, with significant effort, in many specific contexts to better understand physical systems ranging from spin liquids to fractional quantum Hall systems. EHC replaces the insight required to find parent Hamiltonians with an efficient and general approach that can automate their discovery.

While this paper focuses on describing the method and demonstrating its approach through a number of simple illustrative examples, it is important to note that there are many known interesting wave functions that this method could be fruitfully applied to in the future. Examples range from the projective symmetry group (PSG) wave functions \cite{Wen2002,Wang2006,Lu2011,Iqbal2011}, which span a large number of spin liquid phases, to Gutzwiller-projected wave functions, which are heavily used in variational studies of unconventional superconductivity \cite{Gros1988,Gros1989,Paramekanti2001}, to wave functions for fractional Chern insulators \cite{McGreevy2012}. Finding physically realistic parent Hamiltonians for these wave functions could lead to important breakthroughs in spin liquid physics, high temperature superconductivity, and topological phases of matter. There are also a myriad of other potential uses for the EHC framework in fields such as quantum material design, cold-atom quantum simulation, and quantum computing. For example, the EHC framework could help cold-atom experimentalists find Hamiltonians for specific quantum ground states which are constructible within the hardware constraints of their experiment. 

After explaining the method, we discuss three broad applications of EHC. In each application, we discover some unexpected relations between wave functions and the space of Hamiltonians. \\[0.2 cm]
(I) \emph{Hamiltonian Discovery:}  The most straightforward application of EHC is to discover new, simpler, or more experimentally accessible parent Hamiltonians for wave functions without known parent Hamiltonians. To illustrate this procedure, we provide as input to EHC a uniform superposition of frustrated spin configurations and automatically find Hamiltonians with this state as an exact ground state. \\[0.15 cm] 
(II) \emph{State Collision:}  A second application of EHC is to the study of degenerate ground states. Here we introduce a generalized form of EHC, called Degenerate Eigenstate-to-Hamiltonian Construction (DEHC), that receives as input many wave functions and find spaces of Hamiltonians for which those wave functions are degenerate eigenstates. 
DEHC can be used to identify level crossings where two potential phases collide or to identify Hamiltonians with topological degeneracy. We illustrate this approach by colliding the ground states of the Majumdar-Ghosh model \cite{Majumdar1969,Majumdar1970} and the XXZ0 two-leg ladder \cite{Changlani2017}, which are singlet dimer states and projected 3-coloring states, respectively. \\[0.15 cm]
(III) \emph{Phase Expansion: } As a final application, we show how to use EHC to take a known ground state wave function and expand the region of Hamiltonian space over which this wave function is a ground state.  Surprisingly, we discover that many previously known models are in fact special points in large spaces of non-trivial Hamiltonians with identical ground states. We show examples of this procedure by expanding the ground state phase diagram of the XX chain, the Heisenberg chain, the Kitaev chain, and a 2D BdG model.  \\[0.05 cm]

Altogether, in applications (I)-(III), we use seven different types of wave functions as input to EHC and in each case are able to successfully construct new non-trivial parent Hamiltonians.

\section{Method} \label{sec:Method}

\begin{figure*}
\begin{center}
\includegraphics[width=0.9\textwidth]{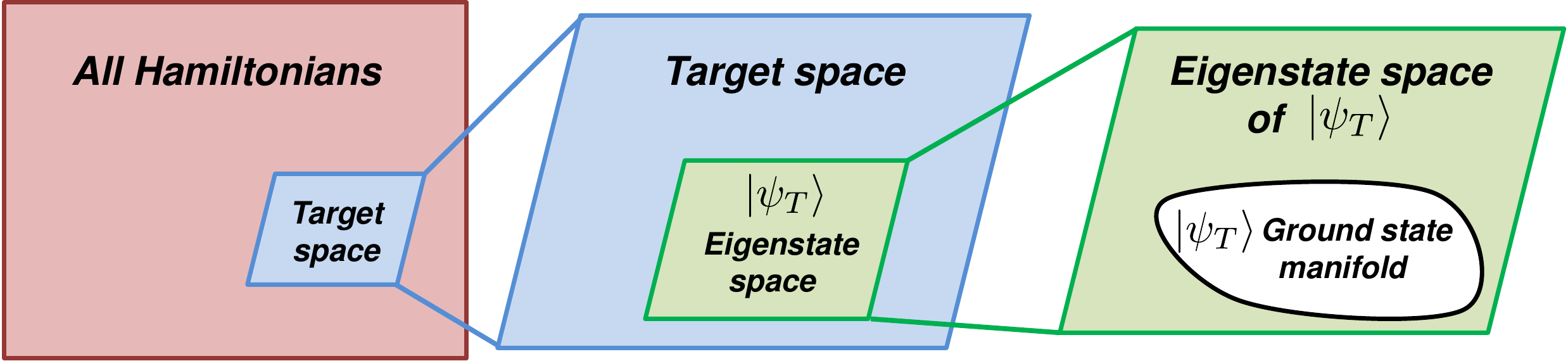}
\end{center}
\caption{All Hamiltonians of a finite-dimensional quantum system form a real vector space of Hermitian operators (shown in red). Eigenstate-to-Hamiltonian Construction (EHC) is performed in a \emph{target space} of Hamiltonians, a physically meaningful subspace of the entire vector space chosen by the user (shown in blue). The output of EHC is the \emph{eigenstate space}, a subspace of the target space consisting of Hamiltonians that contain a target wave function $\ket{\psi_T}$ as an energy eigenstate (shown in green). Using ground state methods, one can further map out the \emph{ground state manifold}, the region in eigenstate space where the target state $\ket{\psi_T}$ is a ground state (shown in white).} \label{fig:HamiltonianSpace}
\end{figure*}

In this section, we introduce our new method, the Eigenstate-to-Hamiltonian Construction (EHC).  EHC takes as input both a target state $\ket{\psi_T}$ and a target space of Hamiltonians $\hat{H}_T$ and produces as output the space of Hamiltonians that contains $\ket{\psi_T}$ as an energy eigenstate, which we call the \emph{eigenstate space} of Hamiltonians. Within the eigenstate space, it is possible for the state $\ket{\psi_T}$ to be a ground state in a particular region, which we call the \emph{ground state manifold}. This hierarchy of Hamiltonian spaces is depicted in Fig.~\ref{fig:HamiltonianSpace}. 

The target space of Hamiltonians $\hat{H}_T$ is a subspace of the vector space of all possible Hamiltonians.  The possible states of a finite system of $N$ quantum degrees of freedom with local dimension $d$, e.g., $d=2$ for $s=1/2$ spins, form a complex vector space of dimension $d^N$. The possible Hamiltonians that can act on this system are all $d^N \times d^N$ Hermitian operators, which form a real vector space of dimension $\left(d^N\right)^2=d^{2N}$.  The target space is a small $d_T$-dimensional \emph{physically meaningful} subspace of Hamiltonian space that we choose when using the EHC method. In particular, we define our target space by choosing a basis of $d_T \ll d^{2N}$ Hermitian operators $\{\hat{h}_a\}_{a=1}^{d_T}$. Defined this way, the target space contains Hamiltonians of the form $\hat{H}_T = \sum_{a=1}^{d_T} J_a \hat{h}_a$ with real $J_a$.  While any set of linearly independent Hermitian operators can be used to define the target space, some natural choices for operators include local one and two-site operators.  

The central tool used in EHC is the \emph{quantum covariance matrix} (QCM), a $d_T \times d_T$ matrix whose matrix elements are given by

\begin{align}
(C_T)_{ab} = \langle\hat{h}_a \hat{h}_b \rangle_T - \langle\hat{h}_a \rangle_T \langle\hat{h}_b \rangle_T \label{eq:CT}
\end{align}
where $\langle \hat{\mathcal{O}} \rangle_T \equiv \bra{\psi_T} \hat{\mathcal{O}}\ket{\psi_T}/\left\langle\psi_T|\psi_T\right\rangle$ and $a,b=1,\ldots,d_T$. The QCM is a quantum-mechanical generalization of a classical covariance matrix, where statistical expectation values of random variables under a probability distribution are replaced by quantum expectation values of Hermitian operators under a wave function. It can be easily shown that $C_T$ is Hermitian and positive semi-definite. Most importantly, $C_T$ can be used to compute the energy variance of the target state for Hamiltonians in the target space:
\begin{align}
\sigma^2_T = \langle \hat{H}_T^2 \rangle_T - \langle \hat{H}_T\rangle_T^2 = \sum_{a=1}^{d_T} \sum_{b=1}^{d_T} J_a (C_T)_{ab} J_b \geq 0. \label{eq:sigmaT}
\end{align}
From Eq.~(\ref{eq:sigmaT}), one can see that an eigenvector of $C_T$ with zero eigenvalue corresponds to a vector of coupling constants $\tilde{J}_a$ and therefore a Hamiltonian $\tilde{H}=\sum_a \tilde{J}_a \hat{h}_a$ with zero energy variance under the target state $\ket{\psi_T}$. Simply by computing the null space of the QCM, we are able to find the eigenstate space of Hamiltonians for $\ket{\psi_T}$.

There are three general cases for the dimensionality of the null space of the QCM. (1)~In the case of a one-dimensional null space, there is a single null vector, and therefore a uniquely specified Hamiltonian \cite{Garrison2015,Villalonga2017,Swingle2014}  in the target space, for which the target state $|\psi_T\rangle$ is an eigenstate. (2)~In the case of a many-dimensional null space, there is a multi-dimensional space of Hamiltonians, which includes any Hamiltonian which can be constructed from a linear combination of the null vectors,  which have $|\psi_T\rangle$ as an eigenstate. The null vectors obtained from numerical decompositions are often in poor representations that are difficult to interpret. To overcome this issue we use an algorithm, described in Ref.~\onlinecite{Qu2014}, which heuristically generates the sparsest basis for the null space. This ensures that each Hamiltonian generated from a basis state in our eigenstate space is constructed from only a small number of distinct Hermitian operators $\hat{h}_a$.  While we find this decomposition fruitful in understanding the resulting Hamiltonians, it is still an important open problem to determine other useful ways of representing the vectors in the null space. (3)~Finally, in the case when the QCM has no null space, the target state $|\psi_T\rangle$ is not an eigenstate of any Hamiltonian within the chosen target space of Hamiltonians.  Nonetheless, the smallest eigenvalues of the QCM still potentially contain useful information. Eigenvectors of the QCM with small eigenvalues correspond to Hamiltonians with small variance under the target state $|\psi_T\rangle$.  This means that the lowest eigenvectors of the QCM represent Hamiltonians under which the target wave function $|\psi_T\rangle$ is ``close'' to an eigenstate. It will be important future work to better understand the implications of this.

EHC is a simple, non-iterative, and remarkably efficient procedure that only requires the computation of $d_T^2+d_T$ expectation values of correlation functions $\langle \hat{h}_a \hat{h}_b \rangle_T$ and observables $\langle \hat{h}_a \rangle_T$. Standard numerical methods for computing such expectation values, such as VMC, DMRG, or exact diagonalization (ED), can be used to evaluate the entries of the QCM. For our specific calculations, we used DMRG, i.e., matrix product state (MPS) methods, and VMC.

When using DMRG, we represent the target state $\ket{\psi_T}$ as a MPS, and the Hermitian operators $\hat{h}_a$  as low bond dimension matrix product operators (MPOs). This allows us to efficiently evaluate $\langle \hat{h}_a \hat{h}_b \rangle_T$ and $\langle \hat{h}_a \rangle_T$ with standard methods, by contracting the MPS $\ket{\psi_T}$ with the $\hat{h}_a \hat{h}_b$ and $\hat{h}_a$ MPOs. We performed our MPS calculations on finite size systems of up to $N=32$ sites and were able to compute all of the entries of the QCM to machine precision.

With VMC, we estimated the expectation values of observables $\hat{\mathcal{O}} \in \{\hat{h}_a\hat{h}_b,\hat{h}_a\}$ under the variational target wave function $\ket{\psi_T}$:
\begin{align*}
\langle\hat{\mathcal{O}}\rangle_T = \frac{\langle \psi_T | \hat{\mathcal{O}} |\psi_T \rangle}{\langle \psi_T | \psi_T \rangle} = \sum_{R} \frac{|\langle \psi_T | R\rangle|^2}{\sum_{R'}|\langle \psi_T | R'\rangle|^2} \frac{\langle R|\hat{\mathcal{O}}|\psi_T\rangle}{\langle R |\psi_T \rangle}
\end{align*}
by sampling configurations $\ket{R}$ from the probability distribution $\propto |\langle \psi_T| R\rangle|^2$ with Metropolis Markov chain Monte Carlo and computing $O(R)\equiv \langle R|\hat{\mathcal{O}}|\psi_T\rangle/\langle R |\psi_T \rangle$ \cite{Becca2017}. The $O(R)$ can be concurrently evaluated during the Markov chain sampling, which means the entries of the QCM have reduced relative statistical noise. Alternatively, with VMC, one can also generate the QCM from a $N_s \times d_T$ sample matrix $M_{sa}=\langle R_s | \hat{h}_a |\psi_T\rangle/\langle R_s | \psi_T \rangle$ for $\{\ket{R_s}\}_{s=1}^{N_s}$ sampled from $\propto |\langle \psi_T |R \rangle|^2$. The sample matrix can be computed more efficiently than the QCM directly, requiring calculating $d_T$ observables per sample instead of $d_T^2$. Moreover, similar to principle component analysis (PCA), one can perform SVD on an appropriately-shifted sample matrix to learn about the eigenvectors of the QCM \footnote{In fact, to learn about the null space of the QCM, one can choose the configurations in the sample matrix arbitrarily.}. 

We empirically found that even though VMC produces a noisy statistical estimate of the QCM, we can still robustly identify properties of the QCM. Interestingly, the dimensionality of the null space has significantly lower statistical noise than the numerical entries of the null vectors.

\begin{figure}
\begin{center}
\includegraphics[width=0.5\textwidth]{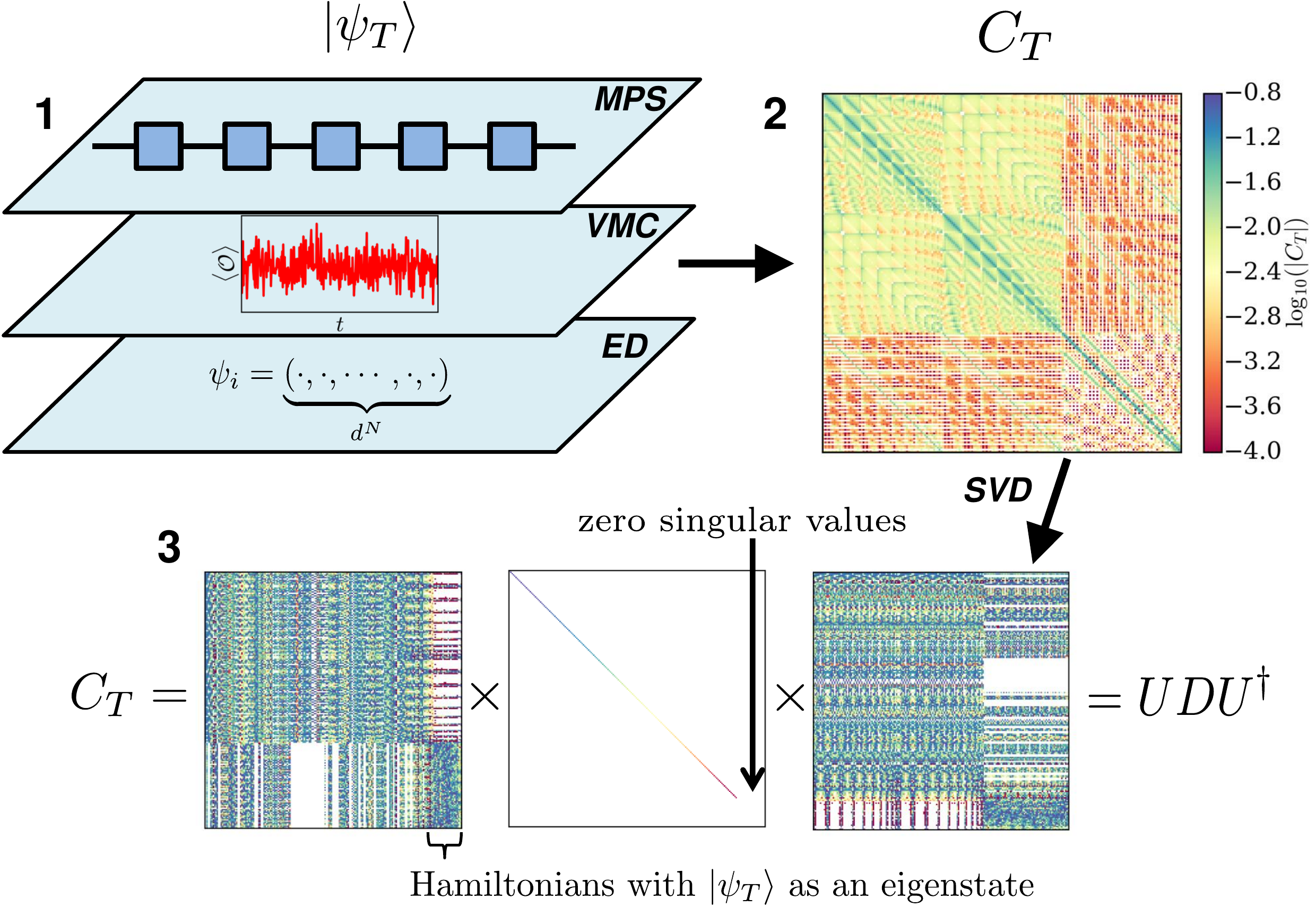}
\end{center}
\caption{The steps of Eigenstate-to-Hamiltonian Construction (EHC). \textbf{1}~Represent the target state $\ket{\psi_T}$ numerically using, for example, MPS, VMC, or ED techniques.
\textbf{2}~Compute the quantum covariance matrix (QCM) $C_T$ given by Eq.~(\ref{eq:CT}). \textbf{3}~Perform a singular value decomposition (SVD) to decompose $C_T$ into $C_T= UDU^\dagger$, with the columns of $U$ representing the singular vectors and the diagonal entries of $D$ representing their corresponding singular values. Identify the null vectors, i.e., the singular vectors with zero singular values. The null vectors correspond to the coupling constants of Hamiltonians with $\ket{\psi_T}$ as an energy eigenstate. The $C_T$ matrix depicted is the QCM for the XX chain ground state used in our phase expansion example.} \label{fig:EHC}
\end{figure}

Finally, in addition to EHC, we developed a generalized form of the method for finding a space of Hamiltonians with multiple target wave functions $\ket{\psi_{T,1}}, \ket{\psi_{T,2}}, \ldots$ as \emph{degenerate} energy eigenstates, called Degenerate Eigenstate-to-Hamiltonian Construction (DEHC), that we discuss in the supplemental material.

\section{Results and discussion}

With a few illustrative examples, we demonstrate three applications of the EHC method -- Hamiltonian discovery, state collision, and phase expansion. Additional examples of phase expansion on the Heisenberg chain and the Kitaev chain are discussed in the supplemental material. A brief summary of our results using the EHC method are shown in Table~\ref{tab:summary}.

\begin{table}
\begin{center}
\begin{tabular}{|c|c||c|c|}
\hline
Target state(s) & $d_T$ & Dim. e.s. space & Dim. g.s. manifold \\
\hline
$\ket{\psi_{UFI}}$ & $111$ & $21$ & $\geq 3$ \\
\hline
$\ket{\psi_{SD}^\pm}$ & $8$ & $4$ & $4$ \\
$\ket{\psi_{P3C}^{m,l}}$ & $8$ & $3$ & $3$ \\
$\ket{\psi_{SD}^\pm}$ \& $\ket{\psi_{P3C}^{m,l}}$ & $8$ & $2$ & $2$ \\
\hline
$\ket{\psi_{XX}}$ & $198$ & $22$ & $\geq 3$ \\
$\ket{\psi_{BCS}}$ & $408$ & $16$ & $\geq 2$ \\

$\ket{\psi_{KC}^\pm}$ & $210$ & $77$ & $\geq 22$ \\
$\ket{\psi_{H}}$ & $198$ & $39$ & $\geq 3$ \\
$\ket{\psi_{SD}^\pm}$ & $198$ & $108$ & $\geq 36$ \\
\hline
\end{tabular}
\end{center}
\caption{A summary of the results of Eigenstate-to-Hamiltonian Construction (EHC) calculations performed in this work. Target states and a target space of Hamiltonians of dimension $d_T$ are provided as input to EHC. The output of EHC is an eigenstate space of Hamiltonians. Ground state methods were used to map out the ground state manifold, but often could only provide a lower bound on the dimensionality of the manifold. The first row is our Hamiltonian discovery result, the next three rows are our state collision results, and the last rows are our phase expansion results. The phase expansion results for the ground states of the Kitaev chain, Heisenberg chain, and Majumdar-Ghosh model, $\ket{\psi_{KC}^{\pm}},\ket{\psi_H},\ket{\psi_{SD}^\pm}$, respectively, were obtained for length $N=12$ chains and are discussed in the supplemental material.} \label{tab:summary}
\end{table}

\subsection{Hamiltonian discovery} \label{sec:HamiltonianDiscovery}

In this section, we investigate a new type of wave function and use the EHC method to construct a parent Hamiltonian for which it is a ground state.

\begin{figure}
\begin{center}
\begin{tabular}{cc}
\includegraphics[width=0.09\textwidth]{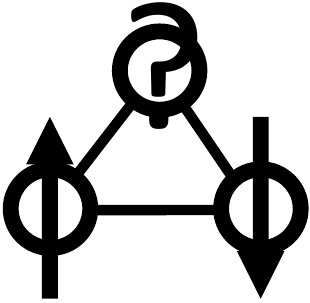} & \includegraphics[width=0.38\textwidth]{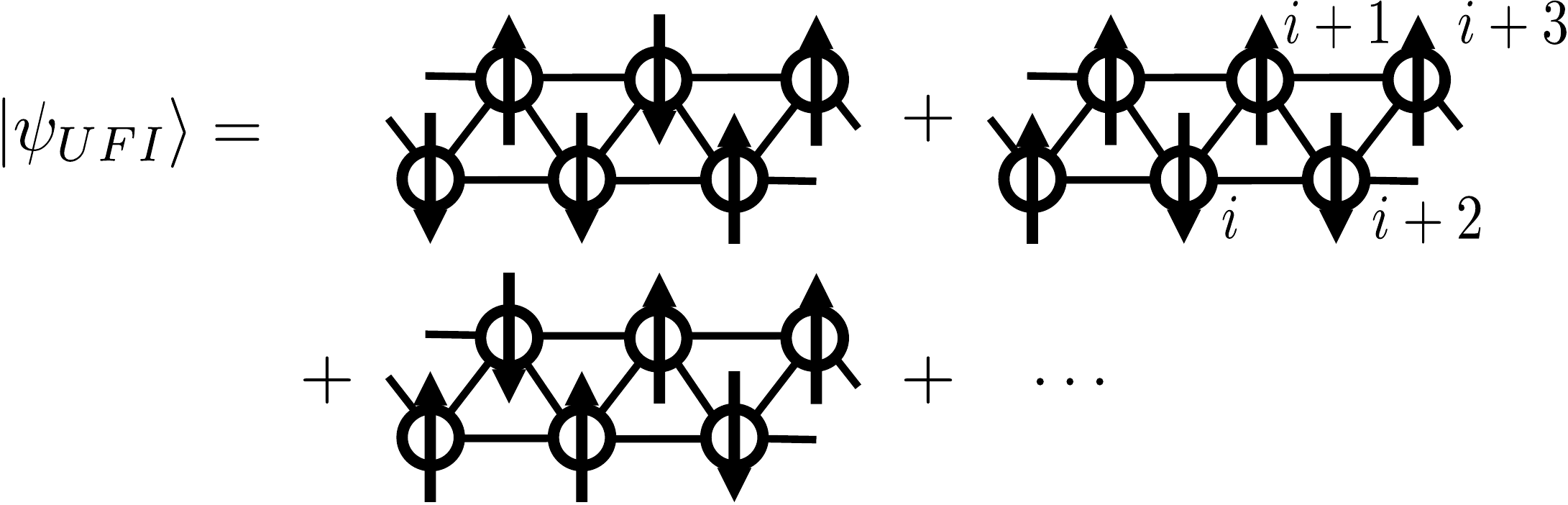} \\
(a) & (b)
\end{tabular}
\end{center}
\caption{(a) For a classical Ising antiferromagnetic triangle, all three bond energies cannot be simultaneously minimized, leading to a six-fold ground state degeneracy. (b) The uniform frustrated Ising (UFI) state is a uniform superposition of the ground states of an antiferromagnetic Ising model on a lattice of triangles. In this case, we consider the UFI state on a triangular two-leg ladder.} \label{fig:UFI}
\end{figure}

We choose a quantum state that is derived from classical magnetically frustrated spin configurations. As shown in Fig.~\ref{fig:UFI}(a), an Ising antiferromagnet on a triangle-tiled lattice exhibits geometric frustration, which results in a large ground state degeneracy. The simplest such model is the antiferromagnetic Ising model on the triangular two-leg ladder: $\hat{H}_I=\sum_{i=1}^N \left( \sigma^z_i \sigma^z_{i+1} + \frac{1}{2} \sigma^z_i \sigma^z_{i+2} \right)$. This model contains combinatorially many ground state spin configurations chosen so that each triangle has exactly two up or two down spins.  We analyze a new type of wave function, called the uniform frustrated Ising (UFI) state $\ket{\psi_{UFI}}$, shown in Fig.~\ref{fig:UFI}(b), which is an equal superposition of the $\hat{H}_I$ ground state spin configurations. We use the UFI state as the target state in EHC and discover Hamiltonians for which it is the ground state. While we do not study the properties of this state, our consideration of it is inspired by other wave functions that are uniform superpositions of ordered states, such as the uniform RVB state \cite{Anderson1973,Anderson1987} and uniform dimer states \cite{Rokhsar1988,Moessner2001,Misguich2002}, which have played important roles in spin liquid physics. We represented our target wave function $\ket{\psi_T}=\ket{\psi_{UFI}}$ numerically as a matrix product state (MPS) for finite periodic ladders of size $N=8,12,16$.

In addition to the target wave function $\ket{\psi_T}$, EHC needs a target space of physically meaningful Hamiltonians $\hat{H}_T=\sum_{a=1}^{d_T} J_a \hat{h}_a$ in which to search for parent Hamiltonians. For our target space of Hamiltonians, we considered a large $d_T=111$ dimensional space of Hamiltonians spanned by periodic, local operators made from products of Pauli matrices on up to $3$ sites separated spatially up to a distance of $3$ sites away on the ladder. Some operators in this target space include:
\begin{align}
\hat{h}_1 \equiv \sum_{i=1}^N \sigma^x_i &\quad\quad \hat{h}_5 \equiv \sum_{i=1}^N \sigma^x_i \sigma^y_{i+1} \nonumber \\
\hat{h}_{28} \equiv \sum_{i=1}^N \sigma^z_i \sigma^x_{i+3} &\quad\quad \hat{h}_{66} \equiv \sum_{i=1}^N \sigma^x_i \sigma^z_{i+1} \sigma^z_{i+3}.
\end{align}

Using the finite MPS representation of $\ket{\psi_{UFI}}$, we computed the $111 \times 111$ quantum covariance matrix (QCM) for the target space defined by the operators $\hat{h}_1,\ldots,\hat{h}_{111}$. We identified $21$ null vectors of the QCM. These correspond to a space of $21$ Hamiltonians with $\ket{\psi_{UFI}}$ as an eigenstate, which includes the Ising model $\hat{H}_I$. Here, we focus on two non-trivial operators in this space
\begin{align*}
\hat{H}_{UFI}^{(1)} &= \sum_{i=1}^N \frac{1}{2}\sigma^z_{i} \sigma^z_{i+2} + \sigma^z_{i} \sigma^z_{i+3} + \sigma^x_{i} \sigma^z_{i+1}\sigma^z_{i+2} - \sigma^z_{i-2} \sigma^x_{i}\sigma^z_{i+1} \\
\hat{H}_{UFI}^{(2)} &= \sum_{i=1}^N \frac{1}{2}\sigma^z_{i} \sigma^z_{i+2} + \sigma^z_{i} \sigma^z_{i+3} + \sigma^z_{i-2} \sigma^z_{i-1}\sigma^x_{i} - \sigma^z_{i-1} \sigma^x_{i}\sigma^z_{i+2}.
\end{align*}
which contain Ising interactions within and between triangles on the ladder as well as off-diagonal three site interactions of the form $\sigma^x \sigma^z \sigma^z$ between triangles. These operators exist in four-site unit cells. We found that the UFI state is an $E=-N/2$ energy eigenstate of these operators. Other operators in the eigenstate space are discussed in the supplemental material.

Using DMRG and ED for ladders of $N=8,12,16,20$ sites, we studied the ground state manifold of the Hamiltonian 
\begin{align}
J\hat{H}_I + J_1 \hat{H}_{UFI}^{(1)} + J_2 \hat{H}_{UFI}^{(2)}. \label{eq:HUFI}
\end{align}
Interestingly, for $J>0$ and $0 < J_1/J=J_2/J < J_c/J$, we found that the UFI state is a ground state of Eq.~(\ref{eq:HUFI}), where $J_c/J$ depends on system size ($J_c/J\approx 0.25,0.20,0.175$ for $N=8,12,16$, respectively). Moreover, for the system sizes studied, we determined empirically that for this range of parameters $\ket{\psi_{UFI}}$ exists in a degenerate ground state manifold containing $5+N/4$ states. Ultimately, our results show that there is a family of quantum models adiabatically connected to the antiferromagnetic Ising two-leg ladder with the UFI state as a ground sate. 

\subsection{State collision} \label{sec:StateCollision}
In this section, we consider many wave functions at the same time and use our inverse method to construct parent Hamiltonians that have all of them as degenerate ground states.

Naively, one might attempt to solve this problem by applying the EHC method repeatedly, once for each wave function, and combining the results. However, this approach would only reveal where in Hamiltonian space the wave functions are simultaneous eigenstates, but not where they are \emph{degenerate} eigenstates. To properly solve this problem, one needs to use the generalization of EHC for degenerate wave functions, DEHC, to find the appropriate eigenstate space. In the following example, we apply DEHC to a triangular two-leg ladder system and search for the eigenstate space where two singlet dimer states and all ``projected 3-coloring states'' are degenerate.      

The singlet dimer states are 
$\ket{\psi_{SD}^\pm}=(\ket{\psi_{1,2}\psi_{3,4}\cdots\psi_{N-1,N}} \pm \ket{\psi_{2,3}\psi_{4,5}\cdots\psi_{N,1}})/\sqrt{2}$ where $\ket{\psi_{i,j}} \equiv (\rket{\uparrow_i \downarrow_j} - \rket{\downarrow_i \uparrow_j})/\sqrt{2}$ is a singlet dimer between sites $i$ and $j$. The states $\ket{\psi_{SD}^\pm}$ are the two degenerate ground states of the periodic Majumdar-Ghosh model \cite{Majumdar1969,Majumdar1970}.

The projected 3-coloring states \cite{Changlani2017} are projected product states of the form 
$\ket{\psi_{P3C}^{m,l}} \equiv  P_{S_z=m} \Big( \bigotimes_{i=1}^N \ket{n_i} \Big) $
where  $\ket{n_i} \equiv (\rket{\uparrow_i} + \omega^{n_i} \rket{\downarrow_i} )/\sqrt{2}$ with $n_i \in \{0,1,2\}$, $P_{S_z=m}$ is a projection onto the $S_z$-sector with magnetization $m$, $\omega \equiv e^{i2\pi/3}$, and each triangle of the two-leg ladder is 3-colored so that $n_i,n_{i+1},n_{i+2}$ are different for every $i$. The parameter $l$ labels the two possible 3-colorings of the two-leg ladder. There are $2N$ linearly independent projected 3-coloring states $\ket{\psi_{P3C}^{m,l}}$.

As input to DEHC, we provide a $d_T=8$ dimensional target space of Hamiltonians spanned by local two-site exchange and Ising interactions on even and odd sites. The first four operators, which act on even sites, are
\begin{align}
\hat{h}_1\equiv\sum_{i=1}^{N/2} \sum_{\rho=x,y} S^\rho_{2i} S^\rho_{2i+1} &\quad\quad \hat{h}_2\equiv\sum_{i=1}^{N/2} S^z_{2i} S^z_{2i+1} \nonumber \\
\hat{h}_3\equiv\sum_{i=1}^{N/2} \sum_{\rho=x,y} S^\rho_{2i} S^\rho_{2i+2} &\quad\quad \hat{h}_4\equiv\sum_{i=1}^{N/2} S^z_{2i} S^z_{2i+2}
\end{align}
where $S^\rho_i=\sigma^\rho_i/2$ are spin-$1/2$ operators. The other four act on odd sites. Also provided as input to DEHC, are all $2N+2$ projected 3-coloring and singlet dimer states.  

From a single DEHC calculation, we found the following two-dimensional space of Hamiltonians for which the singlet dimer states and projected 3-coloring states are degenerate eigenstates 
\begin{align}
&K_1\left(\sum_{i=1}^N\hat{H}_{XXZ0}^{(i,1)} + \frac{1}{2}\sum_{i=1}^N\hat{H}_{XXZ0}^{(i,2)}\right) + K_2\sum_{i=1}^{N}(-1)^i\hat{H}_{XXZ0}^{(i,2)} \label{eq:HIII}
\end{align}
with parameters $K_1$ and $K_2$ defining the space, where $\hat{H}_{XXZ0}^{(i,r)}\equiv S^x_{i}S^x_{i+r} + S^y_{i}S^y_{i+r} - \frac{1}{2} S^z_{i}S^z_{i+r}$ \footnote{In terms of Eq.~(\ref{eq:HI}), this corresponds to $K_1 \equiv J_{xy}, K_2 \equiv \delta_{xy}$ subject to the constraints $J_{z}=-J_{xy}/2,\delta_{z}=-\delta_{xy}/2$. In terms of Eq.~(\ref{eq:HII}), this corresponds to $K_1 \equiv J_{XXZ0}, K_2 \equiv \epsilon_e - J_{XXZ0}/2 = -\epsilon_o + J_{XXZ0}/2$ or $\epsilon_e = K_1/2 + K_2$ and $\epsilon_o=K_1/2 - K_2$.}. This space of Hamiltonians is where the two sets of states ``collide'' and become degenerate with one another.

To better understand the Hamiltonians surrounding this ``collision region,'' we performed two more DEHC calculations, one with only the singlet dimer states as input and one with only the projected 3-coloring states as input. In both cases, we considered the same 8-dimensional target space of Hamiltonians described above.

From one calculation, we found that the singlet dimer states $\ket{\psi_{SD}^\pm}$ are degenerate energy eigenstates of a four-dimensional space of Hamiltonians
\begin{align}
&\sum_{i=1}^N \sum_{\rho=x,y} \left[J_{xy}\left(S^\rho_{i}S^\rho_{i+1} + \frac{1}{2} S^\rho_{i}S^\rho_{i+2}\right) + (-1)^i \delta_{xy} S^\rho_{i} S^\rho_{i+2} \right] \nonumber \\
&+\sum_{i=1}^N \left[J_z \left(S^z_{i}S^z_{i+1} + \frac{1}{2} S^z_{i}S^z_{i+2}\right) + (-1)^i \delta_{z} S^z_{i} S^z_{i+2} \right] \label{eq:HI}
\end{align}
with parameters $J_{xy},J_{z},\delta_{xy},\delta_z$ defining the space.

From the other calculation, we found that the projected 3-coloring states $\ket{\psi_{P3C}^{m,l}}$ are degenerate eigenstates of a three-dimensional space of Hamiltonians
\begin{align}
J_{XXZ0}\sum_{i=1}^{N} \hat{H}_{XXZ0}^{(i,1)} + \epsilon_{e}\sum_{i=1}^{N/2} \hat{H}_{XXZ0}^{(2i,2)} + \epsilon_{o}\sum_{i=1}^{N/2} \hat{H}_{XXZ0}^{(2i+1,2)} \label{eq:HII}
\end{align}
where $J_{XXZ0}, \epsilon_{e},\epsilon_{o}$ are the three parameters defining the space.

\begin{figure}
\begin{center}
\begin{tabular}{c}
\includegraphics[width=0.45\textwidth]{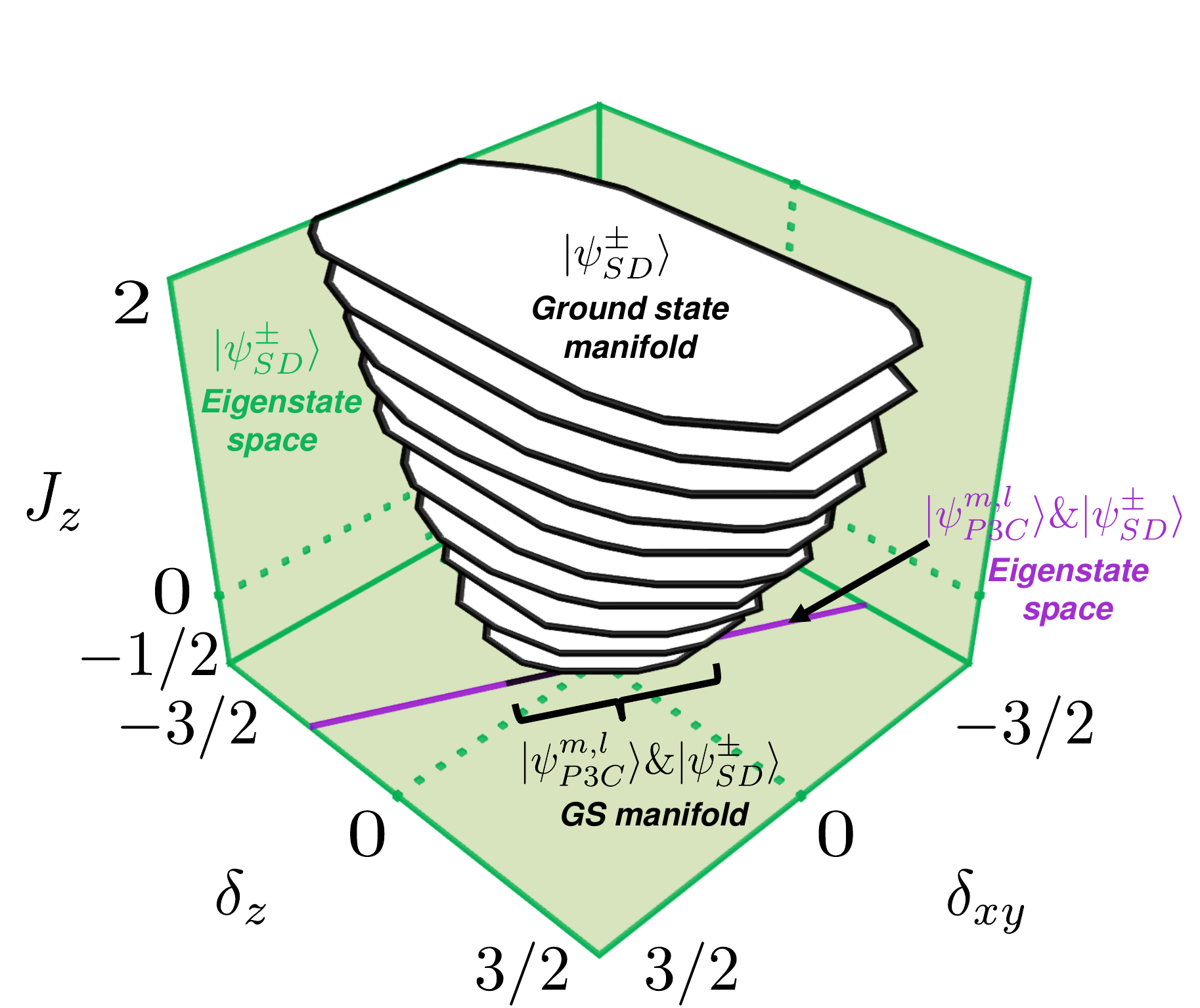} \\
(a) \\
\includegraphics[width=0.45\textwidth]{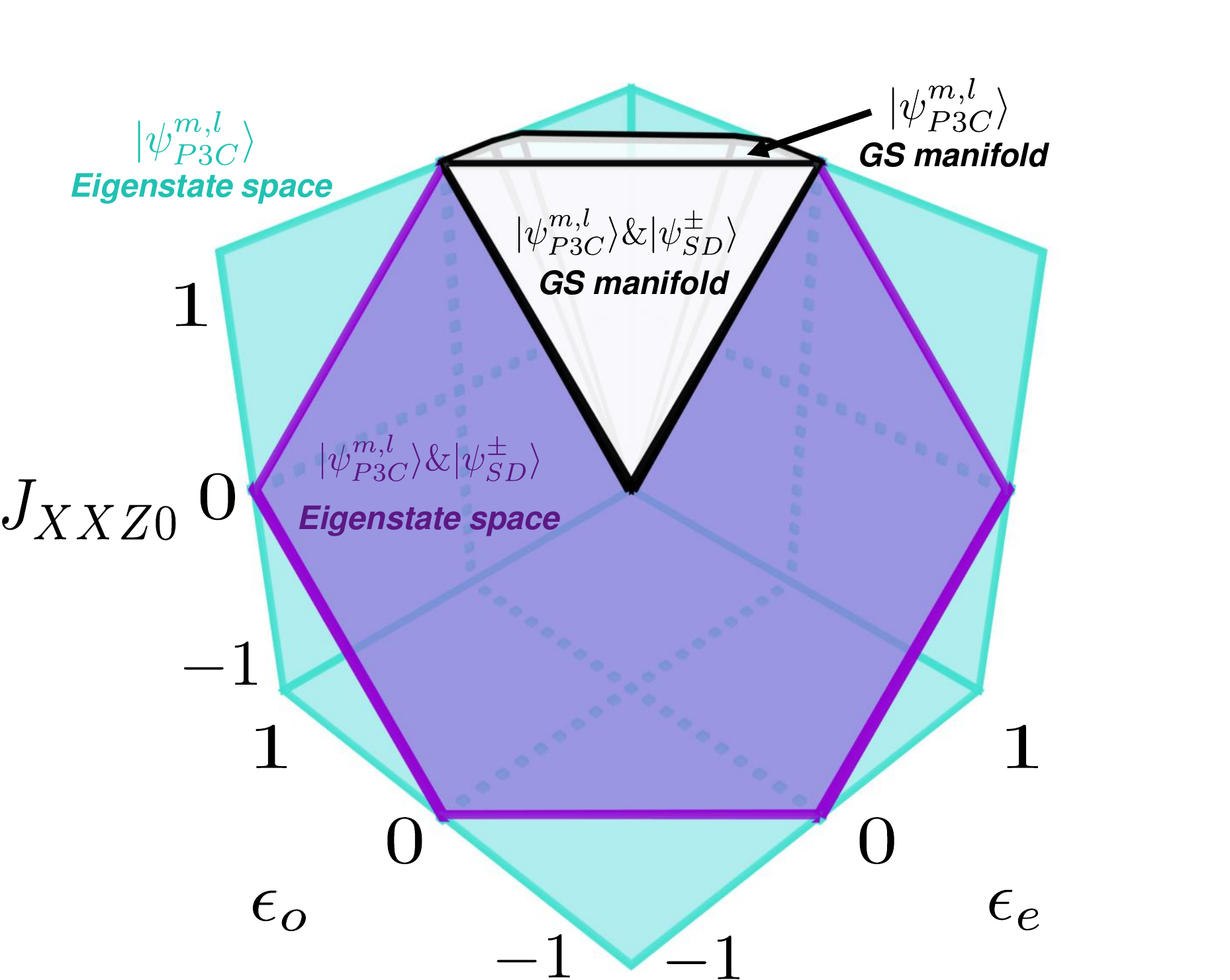} \\
(b)
\end{tabular}
\end{center}
\caption{\textbf{State collision example.}~Collision of the singlet dimer states $|\psi_{SD}^\pm\rangle$ and the projected 3-coloring states $|\psi_{P3C}^{m,l}\rangle$ from (a)  the perspective of the singlet dimer eigenstate space, which is given by Eq.~(\ref{eq:HI}) with $J_{xy}=1$ and is shown in green and (b) the perspective of the projected 3-coloring eigenstate space, which is given by Eq.~(\ref{eq:HII}) and is shown in teal. The space where all of the states are degenerate eigenstates is given by Eq.~(\ref{eq:HIII}) and is shown in purple; it appears as a line in (a) and a plane in (b). The collision region, where all states are degenerate ground states, occurs on the indicated line segment in (a) and in the indicated triangular region in (b).} \label{fig:StateCollision}
\end{figure}

Informed by our inverse method calculations, we could effectively map out the ground state manifolds of the singlet dimer states and the projected 3-coloring states by performing DMRG on the models defined by Eqs.~(\ref{eq:HIII}),~(\ref{eq:HI}),~(\ref{eq:HII}) on finite size ladders of size $N=12,16,32$. Due to the low dimensionality of the Hamiltonian spaces considered in this example, we are able to visualize how the singlet dimer and projected 3-coloring parent Hamiltonians ``collide'' in Hamiltonian space. A visualization of this collision, shown in two different ways, is depicted in Fig.~\ref{fig:StateCollision}. 

Fig.~\ref{fig:StateCollision}(a) shows the ground state manifold of the singlet dimer states $\ket{\psi_{SD}^\pm}$ contained in a three-dimensional projection of their four-dimensional eigenstate space. Fig.~\ref{fig:StateCollision}(b) shows the ground state manifold of the projected 3-coloring states $\ket{\psi_{P3C}^{m,l}}$ contained in their three-dimensional eigenstate space. The combined eigenstate space where $\ket{\psi_{SD}^\pm}$ and $\ket{\psi_{P3C}^{m,l}}$ are degenerate energy eigenstates, is depicted in purple in both Fig.~\ref{fig:StateCollision}(a) and Fig.~\ref{fig:StateCollision}(b). From DMRG and ED, we found that the collision region occurs for the set of parameters $K_1 > 0$  and $-1/2 \leq K_2/K_1 \leq 1/2$ \footnote{In terms of Eq.~(\ref{eq:HI}), this corresponds to $J_{xy}>0,J_z=-J_{xy}/2,\delta_{z}=-\delta_{xy}/2$, and $-1/2 \leq \delta_{xy}/J_{xy} \leq 1/2$. In terms of Eq.~(\ref{eq:HII}), this corresponds to $J_{XXZ0}>0,\epsilon_e/J_{XXZ0}=1 - \epsilon_o/J_{XXZ0}$ and $0 \leq \epsilon_e/J_{XXZ0} \leq 1$.}. The collision region appears as a line segment in Fig.~\ref{fig:StateCollision}(a) and as a triangular region in Fig.~\ref{fig:StateCollision}(b).

Note that the singlet dimer eigenstate space can be constructed from an anisotropic generalization of a known space of ``block operators'' described by Ref.~\onlinecite{Kumar2002}, which we discuss in the supplemental material. Similarly, the projected 3-coloring eigenstate space is largely made up of a known space of triangle-tiled Hamiltonians described by Ref.~\onlinecite{Changlani2017}.

\subsection{Phase expansion} \label{sec:PhaseExpansion}

In this section, we show with two examples, the XX chain and a 2D BDG model, how EHC can be used to expand the zero-temperature phase diagram about a known Hamiltonian $\hat{H}_0$ to find a non-trivial many-dimensional manifold of Hamiltonians with the same ground state wave function as $\hat{H}_0$. 

When phase expanding from $\hat{H}_0$, there are generically two classes of new Hamiltonians one might find: Hamiltonians that commute with $\hat{H}_0$ and those that do not. Adding $\hat{H}_0$-commuting Hamiltonians to $\hat{H}_0$ not only preserves the target state $|\psi_T\rangle$ as an eigenstate, but actually preserve all eigenstates and only acts to shift the eigenvalues. These Hamiltonians can be the result of conserved quantities, such as total $S_z^2$. In our phase expansion results, we, unsurprisingly, find such Hamiltonians. However, surprisingly, we also find non-$\hat{H}_0$-commuting Hamiltonians with $\ket{\psi_T}$ as an eigenstate. Adding such Hamiltonians to $\hat{H}_0$ preserves the target state $|\psi_T\rangle$ as an energy eigenstate while modifying other eigenstates. For example, in the XX chain and Heisenberg chain  (see supplemental material) we find new Hamiltonians that do not commute with their respective $\hat{H}_0$.

The periodic XX chain, $\hat{H}_{XX} = \sum_{n=1}^{N} \left(S^x_n S^x_{n+1} + S^y_n S^y_{n+1}\right)$, has an antiferromagnetic ground state $\ket{\psi_{XX}}$. We represent $\ket{\psi_{XX}}$ as a MPS, which we obtain by performing DMRG on periodic XX chains of length $N=12$.  Because \emph{a priori} we do not know what a possible expanded phase diagram of $\ket{\psi_{XX}}$ might look like, we considered a large target space of Hamiltonians spanned by $d_T=3\binom{N}{2}=3N(N-1)/2=198$ two-site spin operators of the form
\begin{align}
S^x_{i}S^x_{j} \quad\quad S^y_{i}S^y_{j} \quad\quad S^z_{i}S^z_{j} \label{eq:largeTargetSpace}
\end{align}
where $1 \leq i < j \leq N$. Note that these operators are simple and physically reasonable in that they only involve two-site spin interactions, though they are non-local for spins arranged on a chain. One can easily see that the original Hamiltonian $\hat{H}_{XX}$ is contained in this target space.

Using a MPS representation of $\ket{\psi_{XX}}$, we computed the QCM for the target space given by Eq.~(\ref{eq:largeTargetSpace}), which is depicted in Fig.~\ref{fig:EHC}, and found that its null space was spanned by $22$ null vectors for $N=12$, where $12$ were related to total $S_z$ conservation (see supplement for details) and $4$ appeared to be from finite-size effects. The remaining $6$ null vectors corresponded to a space of Hamiltonians
\begin{align}
\hat{H}^{(c,\epsilon)}_{XX}\equiv \sum_{n=1}^N \epsilon^n f^{(c)}(n) \left(S^x_n S^x_{n+1} + \epsilon S^y_n S^y_{n+1}\right) \label{eq:XXEigenstateSpace}
\end{align}
where $\epsilon = \pm 1$ and $f^{(c)}(n)=1,\sin(2\pi n/N),\cos(2\pi n/N)$ for $c=0,1,2$, respectively. These operators correspond to particular types of sinusoidally modulated and anisotropic XX chain interactions. Note that $\ket{\psi_{XX}}$ is a zero energy eigenstate of all of these operators, except for $\hat{H}^{(0,+)}_{XX}=\hat{H}_{XX}$. Also, the four operators $\hat{H}^{(c,\epsilon)}_{XX}$ for $c=1,2$ and $\epsilon=\pm 1$ do not commute with $\hat{H}_{XX}$. In fact, the six operators in Eq.~(\ref{eq:XXEigenstateSpace}) do not commute with one another except in $\epsilon=\pm 1$ pairs, so that $\left[\hat{H}^{(c,+)}_{XX}, \hat{H}^{(c,-)}_{XX}\right]=0$ for all $c$. 

\begin{figure}
\begin{center}
\begin{tabular}{c}
\includegraphics[width=0.5\textwidth]{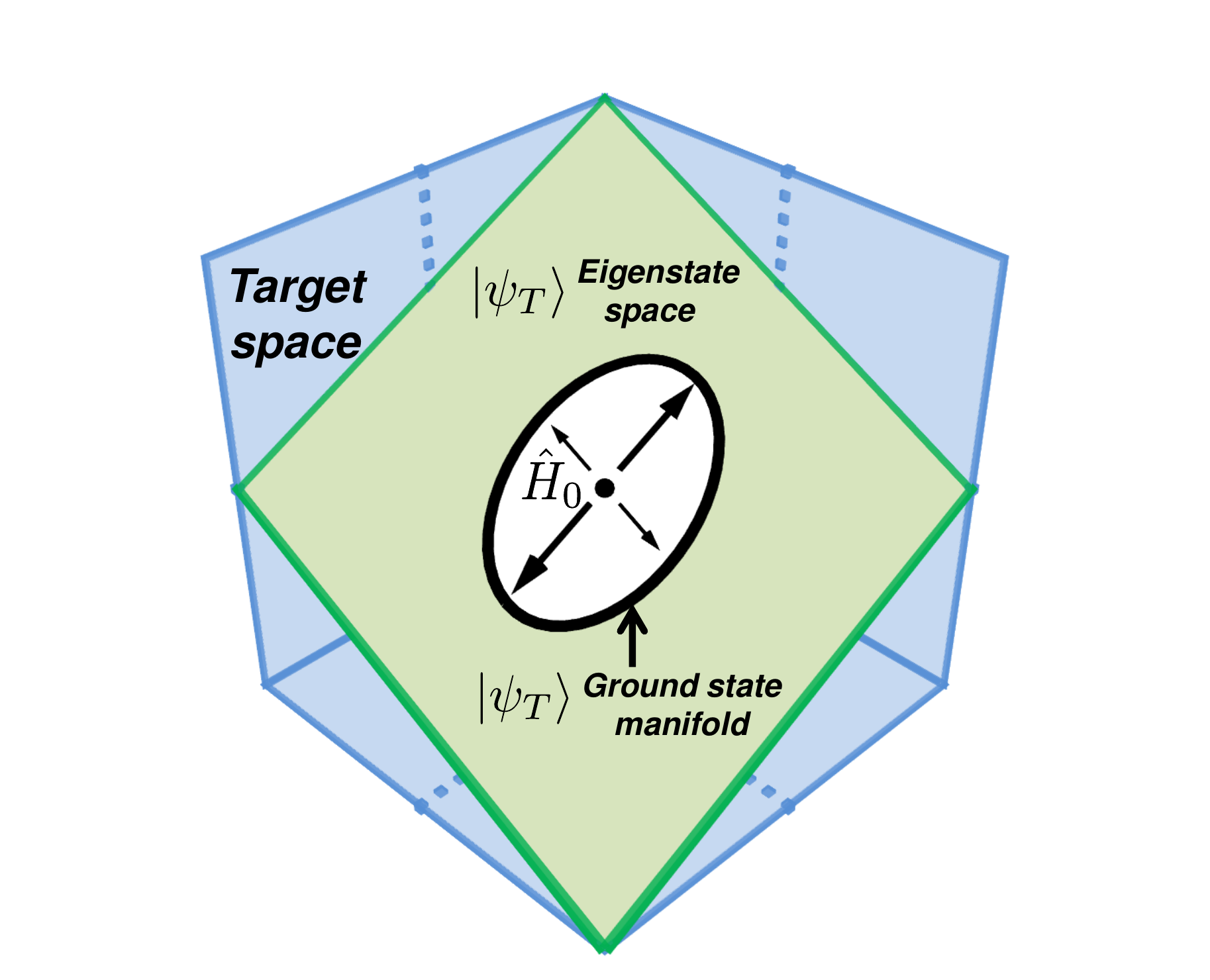} \\
(a) \\ \includegraphics[width=0.45\textwidth]{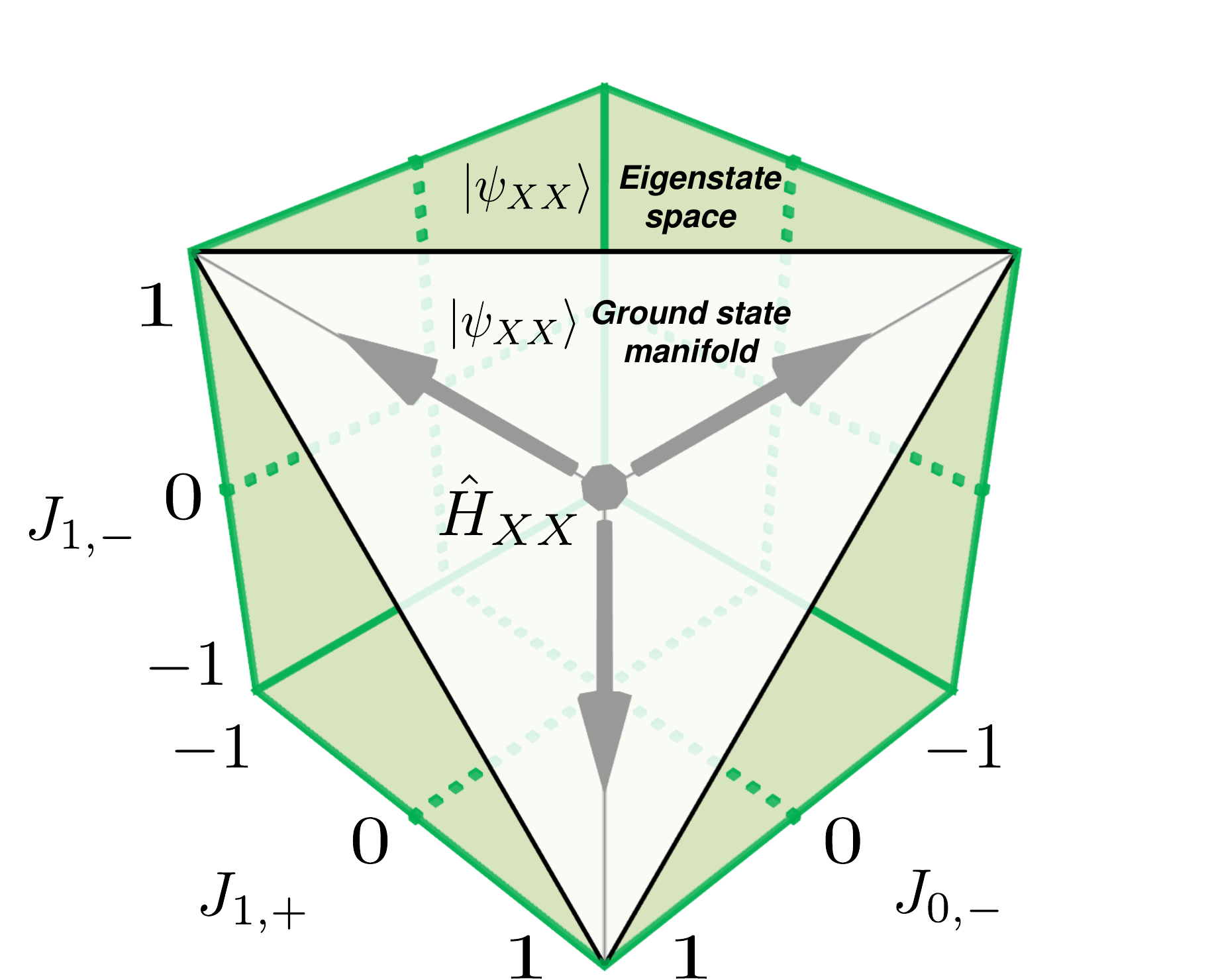} \\
(b)
\end{tabular}
\end{center}
\caption{\textbf{Phase expansion schematic and example.} (a) Schematic representing the expansion of a phase diagram about a known  Hamiltonian $\hat{H_0}$, which has a known ground state $\ket{\psi_T}$. Shown is the target space (blue) provided as input to the EHC method, the eigenstate space (green) of $\ket{\psi_T}$ produced as the output of EHC and the expanded ground state manifold (white) of $\ket{\psi_T}$. (b) Numerical results for the phase expansion about the  Hamiltonian $\hat{H}_{XX}$ with ground state $\ket{\psi_{XX}}$. Shown is a three-dimensional projection (green) of the six-dimensional eigenstate space of $\ket{\psi_{XX}}$, given by Eq.~(\ref{eq:esXX}) with $J_{0,+}=1$ and $J_{2,\pm}=0$ and the ground state manifold of $\ket{\psi_{XX}}$ (white), which almost has the shape of a tetrahedron. For simplicity of visualization, here we plot a tetrahedron, which contains most of the ground state manifold, though ignores a small curved region which extends slightly beyond the tetrahedron.} \label{fig:PhaseExpansion}
\end{figure}

Next, informed by the results of EHC, we mapped out the ground state manifold of $\ket{\psi_{XX}}$ by performing ground state calculations on Hamiltonians in the space described by Eq.~(\ref{eq:XXEigenstateSpace}) \footnote{In this case, the spin-$1/2$ Hamiltonian can be converted into a non-interacting Hamiltonian of spinless fermions with the Jordan-Wigner transformation. The many-body ground state energy can then be efficiently computed by diagonalizing a one-body Hamiltonian.}. We find a highly non-trivial ground state manifold for the Hamiltonian
\begin{align}
\sum_{c=0,1,2} \sum_{\epsilon = \pm 1} J_{c,\epsilon} \hat{H}_{XX}^{(c,\epsilon)} \label{eq:esXX}
\end{align}
defined by the six parameters $J_{0,\pm},J_{1,\pm},J_{2,\pm}$. We analyzed the ground state manifold empirically by considering two and three-dimensional projections of this space subject to the constraint $J_{0,+}=1$. For example, for finite size systems we found the following (approximate) two-dimensional regions where $\ket{\psi_{XX}}$ was the ground state: $|J_{0,-}|+|J_{c,\epsilon}| \lesssim 1$ for $c=1,2$ and $\epsilon=\pm 1$; $|J_{c,-}|+|J_{c,+}| \lesssim 1$ for $c=1,2$; and $(J_{1,\epsilon_1})^2+(J_{2,\epsilon_2})^2 \lesssim 1$ for $\epsilon_1,\epsilon_2=\pm 1$. We also observed an example of a three-dimensional ground state manifold for $\ket{\psi_{XX}}$ with the approximate shape of a tetrahedron, depicted in Fig.~\ref{fig:PhaseExpansion}(b). The tetrahedron-like manifold has endpoints at approximately $(J_{0,-},J_{1,+},J_{1,-})=(-1,-1,-1), (1,1,-1),(1,-1,1),(-1,1,1)$ in coupling constant space. One implication of our results is that the ground state of the XX chain is robust to specific sinusoidally modulated XX-like perturbations. Note that the Hamiltonians found in Eq.~(\ref{eq:esXX}) are related to a mapping discussed in Ref.~\onlinecite{Venuti2010}.

Finally, we discuss an application of the EHC method to a two-dimensional system using variational Monte Carlo (VMC) to calculate the QCM. In this example, we performed phase expansion on the ground state of the following BdG Hamiltonian on an $L\times L$ square lattice
\begin{align*}
\hat{H}_{BdG} &= -\sum_{(x,y),\sigma} (c^\dagger_{(x,y),\sigma}c_{(x+1,y),\sigma} + c^\dagger_{(x,y),\sigma}c_{(x,y+1),\sigma}+ h.c.) \\
&\quad+ \sum_{(x,y)} (c_{(x,y),\uparrow}c_{(x,y),\downarrow} + h.c.) \label{eq:HBdG}
\end{align*}
where $(x,y)$ indicates the coordinates of a site in the lattice. This model is the parent Hamiltonian of the s-wave BCS wave function $\ket{\psi_{BCS}}=\prod_k (u_k + v_k c^\dagger_{k\uparrow}c^\dagger_{-k\downarrow})\ket{0}$ with BCS parameters $u_k,v_k$ defined in the standard way and $\Delta_k=\Delta=t=1,\mu=0$.

The target space provided as input to EHC was spanned by all possible one and two-site operators of the form
\begin{gather*}
\sum_{\sigma}n_{(x,y)\sigma},\quad n_{(x,y)\uparrow}n_{(x,y)\downarrow}, \quad \sum_{\sigma}(c^\dagger_{(x,y)\sigma} c_{(x',y')\sigma} + h.c.), \\ \quad (c_{(x,y)\uparrow} c_{(x',y')\downarrow} + h.c.), \quad \sum_{\sigma,\sigma'} n_{(x,y)\sigma} n_{(x',y')\sigma'}.
\end{gather*}
In our calculations, we considered an $N=4\times 4=16$ site system, which made the dimension of this target space $d_T=408$.

Using VMC, we numerically estimated the QCM for the $N=16$ site BCS state in this target space. The eigenstate space produced by the EHC method contained $16$ operators. One interesting Hamiltonian in this space is the staggered s-wave pairing operator
\begin{align}
\hat{H}_{s} &= \sum_{(x,y)} (-1)^{x+y}(c_{(x,y),\uparrow}c_{(x,y),\downarrow} + h.c.).
\end{align}
The s-wave BCS state $\ket{\psi_{BCS}}$ is a zero energy eigenstate of this operator. Numerically, we determined that $\ket{\psi_{BCS}}$ is actually the ground state of the phase expanded model $t\hat{H}_{BdG}+\Delta_s\hat{H}_{s}$ for $t>0$ and $-1 \leq \Delta_s/t \leq 1$, even for large system sizes.  An alternative approach for constructing parent Hamiltonians from BCS ground states is given in Ref.~\onlinecite{Wang2017}.

\begin{figure*}
\begin{center}
\includegraphics[width=0.75\textwidth]{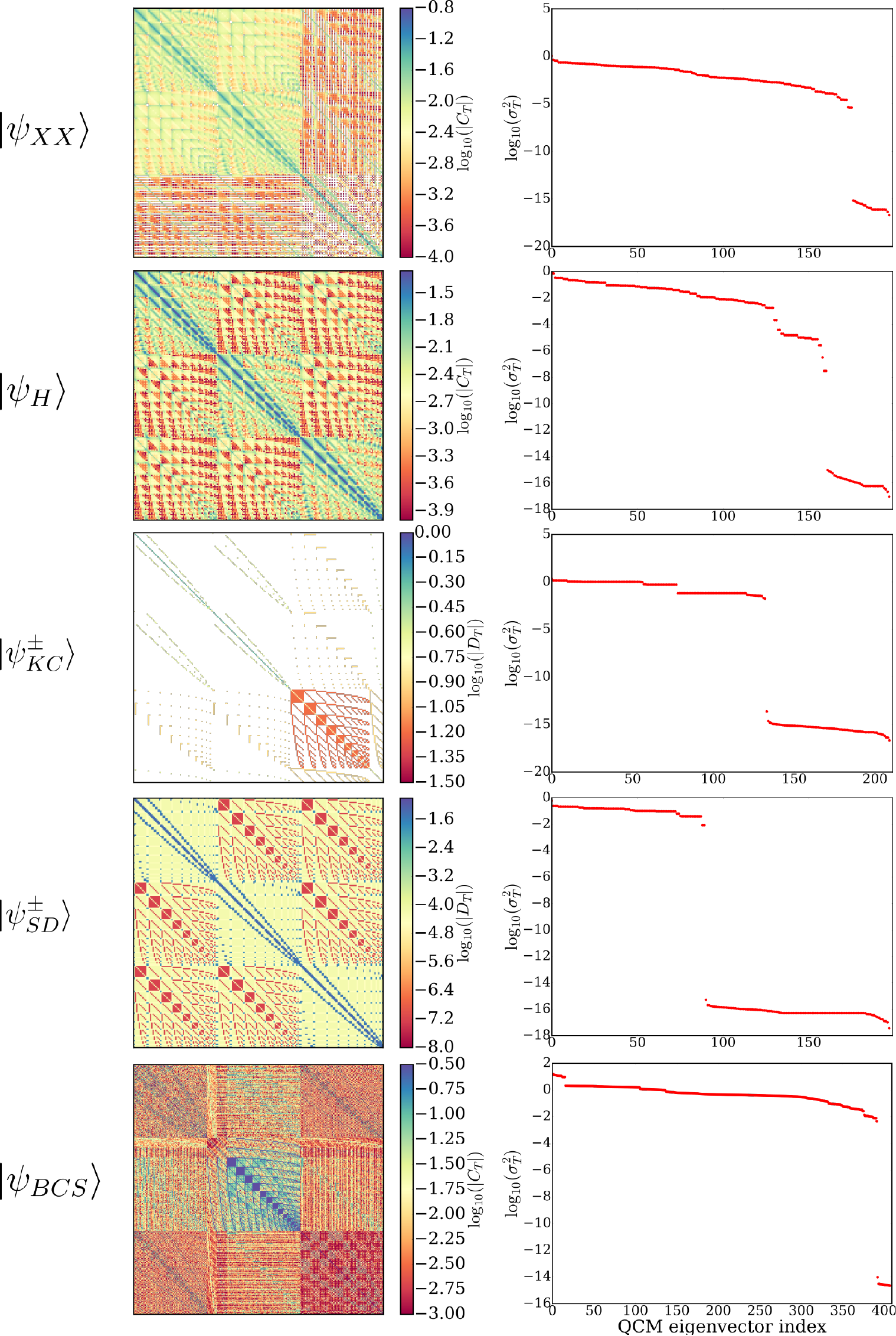}
\end{center}
\caption{A summary of phase expansion results obtained in this work. In each row, we show the target state(s) provided as input to EHC (and DEHC), the QCMs $C_T$ (and DQCMs $D_T$, see supplement) we calculated, and their spectra on a log scale. The eigenvectors of a QCM (or DQCM) are vectors of coupling constants that correspond to Hamiltonians with variance $\sigma^2_T$ under the target state $\ket{\psi_T}$. The QCM for $\ket{\psi_{BCS}}$ in the bottom row was statistically estimated using variational Monte Carlo, while the others were computed with matrix product states.} \label{fig:QCMs}
\end{figure*}

Other examples of phase expansion for the Kitaev chain, Heisenberg chain, and Majumdar-Ghosh model are discussed in the supplemental material. A visualization of the QCMs computed in our phase expansion results and their spectra are shown in Fig.~\ref{fig:QCMs}. Note that for each QCM considered, there are many eigenvalues that are zero to numerical precision, which are separated by many orders of magnitude from the non-zero eigenvalues. We note that for frustration-free models, such as the Kitaev chain and Majumdar-Ghosh model, we found eigenstate spaces that were much higher-dimensional than for the other models we considered.

\section{Summary}
We have developed the Eigenstate-to-Hamiltonian Construction (EHC) which is an efficient inverse method that can be used to produce spaces of physically meaningful parent Hamiltonians from a wave function.   Analogous to variational wave function approaches to the forward problem, EHC is a variational Hamiltonian approach that finds parent Hamiltonians from a class of models. We anticipate that it will play a similarly important role in strongly-correlated physics.

The key to the EHC method is computing the quantum covariance matrix (QCM) (see Eq.~(\ref{eq:CT})), from which parent Hamiltonians can be found. Even though for the examples presented in this work we computed the QCM using VMC and DMRG, one can certainly compute the QCM using various other analytical, numerical, and experimental approaches.  For example, one can compute the QCM in the context of sign-problem free Hamiltonians using quantum Monte Carlo.  

We have described some sample applications of EHC in which we revealed some interesting and unforeseen structure of Hamiltonian space.  We demonstrated how to find new types of Hamiltonians with EHC by automatically constructing parent Hamiltonians for a uniform superposition of frustrated Ising spin configurations. The discovered parent Hamiltonians are non-trivial quantum models that are adiabatically connected to the degenerate ground state manifold of the classical Ising antiferromagnet. This example clearly illustrates how the EHC method can quickly and with minimal theoretical ingenuity produce parent Hamiltonians that might otherwise take significant effort or insight to discover.

We also showed how the degenerate version of EHC, DEHC, can find regions of Hamiltonian space where many wave functions are degenerate, which allows one to automatically identify level-crossings or topological degeneracies between different states. We demonstrated this by finding the space where singlet dimer states and 3-coloring states ``collide,'' resulting in a highly degenerate ground state manifold corresponding to a first-order quantum phase transition. 

Finally we showed how to use EHC to expand the phase diagram of known model Hamiltonians, such as the XX chain, the Kitaev chain, the Heisenberg chain, and a 2D s-wave BdG model. In doing so, we showed that these specific models are actually special points in Hamiltonian space and that their ground states are shared with surrounding Hamiltonians in large, non-trivial regions in this space.   

\section{Conclusions}

The EHC approach fits into a broader class of techniques, such as machine learning approaches, for automating physical understanding that previously required significant insight.  Moreover, given the relation between the QCM and covariance matrices used in statistics, data science, and machine learning, one might expect methods developed in those contexts, such as principal component analysis, to be applicable to EHC and quantum systems.  

The standard approach to condensed matter physics is to take a Hamiltonian and determine emergent properties represented by it ground state(s). Historically this has been difficult because of the exponential computational complexity in determining the exact ground state wave function. In this work, we invert this approach, demonstrating a new approach to condensed matter physics. Starting with wave functions with desired properties, we find Hamiltonians with these ground states in time quadratic in the dimension of the local Hamiltonian space explored by EHC. This new perspective asks us to consider more broadly the structure of the larger phase space of physically meaningful Hamiltonians.  

EHC is a general tool that allows both theorists and experimentalist to construct Hamiltonians that have interesting physics or targeted properties in their ground states. Example uses might include targeting ground states in cold-atom systems as well as applications to spin liquids, fractional quantum Hall physics, unconventional superconductivity, many-body localization, frustrated magnetism, and continuum ab-initio approaches. EHC is a key step toward the long-term goal of material design of strongly correlated materials.  

\textbf{Note Added: } In the final stages of preparing this manuscript, a post to the Journal Club for Condensed Matter Physics \cite{Vishwanath2018} brought to our attention the existence of a recent preprint \cite{Ranard2017} that independently developed a similar approach based on the quantum covariance matrix.   

\section{Acknowledgements}
Our calculations involving matrix product states were done using the ITensor library \cite{itensor}. This project has received support under SciDAC
grant DE-FG02-12ER46875.  This research is part of the Blue
Waters sustained-petascale computing project, which is
supported by the National Science Foundation (awards
OCI-0725070 and ACI-1238993) and the state of Illinois.
Blue Waters is a joint effort of the University of Illinois
at Urbana-Champaign and its National Center for Supercomputing
Applications. E.C. acknowledges a graduate student travel award to attend the 2017 APS March Meeting where this work was originally presented. 

\bibliography{refs}

\end{document}